\documentclass[a4paper]{article}
\usepackage{a4,amsmath,cite}
\usepackage[dvips]{graphicx}
\usepackage{dcolumn}
\usepackage{caption}

\begin{document}

\large
\title{\textbf{Comparison theorems for the
position-dependent mass Schr\"odinger equation}}

\author{D.A.
Kulikov\thanks{kulikov\_d\_a@yahoo.com, kulikov@dsu.dp.ua},
\\
{\sl Theoretical Physics Department, FFEKS, Dniepropetrovsk
National
University }\\
{\sl 72 Gagarin avenue, Dniepropetrovsk 49010, Ukraine} }

\date{}
\maketitle

\begin{abstract}
The following comparison rules for the discrete spectrum of the
position-dependent mass (PDM) Schr\"odinger equation are
established. (i) If a constant mass $m_0$ and a PDM
$m(\mathbf{x})$ are ordered everywhere, that is either $m_0\leq
m(\mathbf{x})$ or $m_0\geq m(\mathbf{x})$, then the corresponding
eigenvalues of the constant-mass Hamiltonian and of the PDM
Hamiltonian with the same potential and the BenDaniel-Duke
ambiguity parameters are ordered. (ii) The corresponding
eigenvalues of PDM Hamiltonians with the different sets of
ambiguity parameters are ordered if $\nabla^2 (1/m(\mathbf{x}))$
has a definite sign. We prove these statements by using the
Hellmann-Feynman theorem and offer examples of their application.
\end{abstract}


\section{Introduction}
\label{part1}

Last few decades, quantum mechanical systems with
position-dependent mass (PDM) have received considerable
attention. The interest stems mainly from the relevance of the PDM
background for describing the physics of compositionally graded
crystals \cite{bastard,young} and semiconductor nano devices
\cite{herling,peter,khordad}. These applications have stimulated
the study of the various theoretical aspects of the PDM
Schr\"odinger equation; in particular, its exact solvability
\cite{dekar,alhaidari,shihai}, shape invariance \cite{bagchi_si},
supersymmetry and intertwining properties
\cite{tanaka,ganguly,midya}, point canonical transformation
\cite{sever,kraenkel}, iterative solution \cite{koc2010}, and
relation to theories in curved spaces \cite{tkachuk} have been
examined.

However, it is known that the PDM Schr\"odinger equation suffers
from ambiguity in operator ordering, caused by the non-vanishing
commutator of the momentum operator and the PDM. The PDM
Hamiltonians with different ambiguity parameters have been
proposed \cite{roos,bd,lk,gora}, but none of them can be preferred
according to the existing reliability tests
\cite{morrow,mustafa1,almeida}. Therefore the attempts are made to
settle the issue by fitting the calculated binding energies to the
experimental data \cite{cavalcante,smagley}.

For generelizing such findings and obtaining additional
information, one needs some tools to compare the energy
eigenvalues predicted by the different PDM Hamiltonians. Within
the constant-mass framework, a convenient tool is provided by the
so-called comparison theorems \cite{hall_refine,hall_gen,semay}.
For example, the elementary comparison theorem
\cite{hall_refine,semay} states that if two real potentials are
ordered, $V^{(1)}\leq V^{(2)}$, then each corresponding pair of
eigenvalues is ordered, $E^{(1)}\leq E^{(2)}$.

The purpose of this paper is to establish the comparison theorems
that confront the energy eigenvalues of the constant-mass and PDM
Schr\"odinger equations, as well as the energy eigenvalues of the
PDM problems with different ambiguity parameters. Our presentation
is based on the Hellmann-Feynman theorem \cite{Hell35} and makes
use of the ideas developed for the constant-mass case
\cite{semay,hall_rel}.

The plan of the paper is as follows. In Section \ref{part2} we
introduce the PDM Hamiltonians and recall the Hellmann-Feynman
theorem. In Section \ref{part3} the comparison theorems on the PDM
background are formulated and proved. In Section \ref{part4} we
apply these theorems to two PDM problems of current interest.
Finally, our conclusions are summarized in Section \ref{part5}.

\section{Preliminaries}
\label{part2}

For the PDM Schr\"odinger equation the most general form of the
Hamiltonian is given by \cite{roos}
\begin{equation}\label{eq1}
H_{\mathrm{pdm}}=  - \frac{1}{4} \left( m(\mathbf{x})^{\alpha}
\nabla m(\mathbf{x})^{\beta} \nabla m(\mathbf{x})^{\gamma} +
m(\mathbf{x})^{\gamma} \nabla m(\mathbf{x})^{\beta} \nabla
m(\mathbf{x})^{\alpha} \right)+ V (\mathbf{x})
\end{equation}
where $\alpha$, $\beta$, $\gamma$ are the ambiguity parameters
($\alpha + \beta + \gamma = -1$) and the units with $\hbar=1$ are
used. In this paper we will adopt the sets of the ambiguity
parameter values suggested by BenDaniel and Duke \cite{bd}
($\alpha=\gamma=0$, $\beta=-1$), Li and Kuhn \cite{lk}
($\alpha=\beta=-1/2$, $\gamma=0$), and Gora and Williams
\cite{gora} ($\alpha=-1$, $\beta=\gamma=0$).

The methods we are going to apply are valid for arbitrary
dimension $N$. We suppose that the Hamiltonian operators have
domains ${\cal{D}}(H)\subset L^2(R^N)$, they are bounded below,
essentially self-adjoint, and have at least one discrete
eigenvalue at the bottom of the spectrum.

To derive our main results, we need the Hellmann-Feynman theorem
\cite{Hell35}.  This theorem states that if the Hamiltonian of a
system is $H(a)$, where $a$ is a parameter, and the eigenvalue
equation for a bound state is $H(a)|a\rangle = E(a)|a\rangle$,
where $E(a)$ is the energy and $|a\rangle$ the normalized
associated eigenstate, then
\begin{equation}
\label{eq2} \frac{\partial E(a)}{\partial a} = \left\langle a
\left|\frac{\partial H(a)}{\partial a}\right|a\right\rangle.
\end{equation}
Note that the proof relies on the self-adjointness of $H(a)$ and
does not change for PDM Hamiltonians.

\section{Comparison theorems}
\label{part3}

First, let us formulate the theorem that confronts the energy
eigenvalues of the constant-mass and BenDaniel-Duke PDM
Hamiltonians with the same potentials.

\hfil

\noindent {\bf Theorem 1}

\noindent Suppose that the Hamiltonian
\begin{equation}\label{eq3}
H^{(0)}=  - \frac{1}{2m_0}\nabla^2  + V(\mathbf{x})
\end{equation}
with a real potential $V(\mathbf{x})$ and a constant-mass $m_0$
has discrete eigenvalues $E_{\{n\}}^{(0)}$ characterized by a set
of quantum numbers ${\{n\}}$. Then the corresponding eigenvalues
$E_{\{n\}}^{(\mathrm{BD})}$ of the BenDaniel-Duke PDM Hamiltonian
\begin{equation}\label{eq4}
H^{(\mathrm{BD})}=- \frac{1}{2} \nabla \frac{1}{m(\mathbf{x})}
\nabla + V(\mathbf{x})
\end{equation}
satisfy
\begin{eqnarray}
\label{eq5a} && E^{(0)}_{\{n\}}\leq E^{(\mathrm{BD})}_{\{n\}}
\qquad \mathrm{if} \qquad \forall\,\mathbf{x} \quad
0<m(\mathbf{x}) \leq
m_0,  \\
\label{eq5b} && E^{(0)}_{\{n\}}\geq E^{(\mathrm{BD})}_{\{n\}}
\qquad \mathrm{if} \qquad \forall\,\mathbf{x} \quad m(\mathbf{x})
\geq m_0,
\end{eqnarray}
provided that these eigenvalues exist.


\noindent {\bf Proof:} Define the Hamiltonian
\begin{equation}\label{eq6}
H(a)=(1-a)\,H^{(0)}+a\,H^{(\mathrm{BD})},
\end{equation}
which turns into $H^{(0)}$ and $H^{(\mathrm{BD})}$ when $a=0$ and
$a=1$, respectively. Assume that  $H(a)$ possesses well defined
eigenvalues $E_{\{n\}}(a)$, for $0 \leq a \leq 1$, and the
normalized associated eigenfunctions in the coordinate
representation are $\psi_{\{n\}}(\mathbf{x};a)$.

Applying the Hellmann-Feynman theorem (\ref{eq2}), we get
\begin{equation}\label{eq65}
\frac{\partial E_{\{n\}}(a)}{\partial a}=
\int\psi_{\{n\}}^{*}(\mathbf{x};a) \left( \frac{1}{2m_0}\nabla^2 -
\frac{1}{2} \nabla \frac{1}{m(\mathbf{x})} \nabla \right)
\psi_{\{n\}}(\mathbf{x};a)d\mathbf{x}
\end{equation}
where the integration is performed over the whole space and the
asterisk denotes complex conjugation.

Integrating by parts and taking into account that
$\psi_{\{n\}}(\mathbf{x};a)$ and
$\nabla\psi_{\{n\}}(\mathbf{x};a)$ must vanish at infinity, we
obtain
\begin{equation}\label{eq7}
\frac{\partial E_{\{n\}}(a)}{\partial a}=
\frac{1}{2}\int\left(\frac{1}{m(\mathbf{x})}-\frac{1}{m_0}\right)|\nabla\psi_{\{n\}}(\mathbf{x};a)|^2
d\mathbf{x}.
\end{equation}
It is a positive (negative) number if $0<m(\mathbf{x})\leq m_0$
($m(\mathbf{x})\geq m_0$) for all $\mathbf{x}$, so that
$E_{\{n\}}(a)$ is an increasing (decreasing) function of $a$. For
definiteness, let $0<m(\mathbf{x})\leq m_0$. Then it follows
immediately
\begin{equation}\label{eq8}
E_{\{n\}}(0)=E_{\{n\}}^{(0)}\leq
E^{(\mathrm{BD})}_{\{n\}}=E_{\{n\}}(1)
\end{equation}
that completes the proof. Note that an alternative proof can be
given by applying the variational characterization \cite{reed} of
the discrete part of the Schr\"odinger spectrum.


It is now tempting to compare the eigenvalues of the constant-mass
Hamiltonian with those of PDM Hamiltonians other than the
BenDaniel-Duke one. However, in that case we encounter an obstacle
that becomes clear if we first find out how the eigenvalues of
different PDM Hamiltonians are ordered. This is done in the
following theorem.

\hfil

\noindent {\bf Theorem 2}

\noindent The discrete eigenvalues $E_{\{n\}}^{(\mathrm{BD})}$,
$E_{\{n\}}^{(\mathrm{LK})}$ and $E_{\{n\}}^{(\mathrm{GW})}$ of the
BenDaniel-Duke, Li-Kuhn and Gora-Williams PDM Hamiltonians
\begin{eqnarray}
\label{eq12a} && H^{(\mathrm{BD})}=  - \frac{1}{2} \nabla
\frac{1}{m(\mathbf{x})} \nabla +
V(\mathbf{x}), \\
\label{eq12b} && H^{(\mathrm{LK})}=  - \frac{1}{4}\left(
\frac{1}{\sqrt{m(\mathbf{x})}}\nabla
\frac{1}{\sqrt{m(\mathbf{x})}}\nabla+ \nabla
\frac{1}{\sqrt{m(\mathbf{x})}}\nabla\frac{1}{\sqrt{m(\mathbf{x})}} \right)+ V(\mathbf{x}),  \\
\label{eq12c} && H^{(\mathrm{GW})}=  - \frac{1}{4}\left(
\frac{1}{m(\mathbf{x})}\nabla^2+ \nabla^2\frac{1}{m(\mathbf{x})}
\right) + V(\mathbf{x})
\end{eqnarray}
satisfy
\begin{eqnarray}
\label{eq13a} && E^{(\mathrm{BD})}_{\{n\}}<
E^{(\mathrm{LK})}_{\{n\}}< E^{(\mathrm{GW})}_{\{n\}} \qquad
\mathrm{if} \qquad \forall\,\mathbf{x} \quad
\nabla^2\left(\frac{1}{m(\mathbf{x})}\right) < 0,  \\
\label{eq13b} && E^{(\mathrm{BD})}_{\{n\}}>
E^{(\mathrm{LK})}_{\{n\}}> E^{(\mathrm{GW})}_{\{n\}} \qquad
\mathrm{if} \qquad \forall\,\mathbf{x} \quad
\nabla^2\left(\frac{1}{m(\mathbf{x})}\right) > 0,
\end{eqnarray}
provided that these eigenvalues exist.

\noindent {\bf Proof:} Let us prove the inequalities for
$E^{(\mathrm{BD})}_{\{n\}}$ and $E^{(\mathrm{LK})}_{\{n\}}$. We
define the parameter-dependent Hamiltonian $H(a)$ by
\begin{equation}\label{eq14}
H(a)=(1-a)\,H^{(\mathrm{BD})}+a\,H^{(\mathrm{LK})}
\end{equation}
and make use of the Hellmann-Feynman theorem (\ref{eq2}), to
obtain
\begin{eqnarray}\label{eq15}
\frac{\partial E_{\{n\}}(a)}{\partial a}=
\int\psi_{\{n\}}^{*}(\mathbf{x};a) \left[ \frac{1}{2} \nabla
\frac{1}{m(\mathbf{x})} \nabla \right. \nonumber\\
\left.- \frac{1}{4}\left( \frac{1}{\sqrt{m(\mathbf{x})}}\nabla
\frac{1}{\sqrt{m(\mathbf{x})}}\nabla+ \nabla
\frac{1}{\sqrt{m(\mathbf{x})}}\nabla\frac{1}{\sqrt{m(\mathbf{x})}}
\right) \right] \psi_{\{n\}}(\mathbf{x};a)d\mathbf{x}.
\end{eqnarray}

Integration by parts yields
\begin{eqnarray}\label{eq16}
&&\frac{\partial E_{\{n\}}(a)}{\partial
a}=-\frac{1}{2}\int\frac{1}{m(\mathbf{x})}|\nabla\psi_{\{n\}}(\mathbf{x};a)|^2
d\mathbf{x}\nonumber \\
&&+\frac{1}{4}\int
\nabla\left(\frac{1}{\sqrt{m(\mathbf{x})}}\psi_{\{n\}}^{*}(\mathbf{x};a)\right)\cdot
\frac{1}{\sqrt{m(\mathbf{x})}}\nabla \psi_{\{n\}}(\mathbf{x};a)d\mathbf{x}\nonumber \\
&&+\frac{1}{4}\int
\nabla\left(\frac{1}{\sqrt{m(\mathbf{x})}}\psi_{\{n\}}(\mathbf{x};a)\right)\cdot
\frac{1}{\sqrt{m(\mathbf{x})}}\nabla \psi_{\{n\}}^{*}(\mathbf{x};a)d\mathbf{x}\nonumber \\
&& =-\frac{1}{8}\int\frac{1}{m(\mathbf{x})^2}\left(\nabla
m(\mathbf{x})\right)\cdot\nabla|\psi_{\{n\}}(\mathbf{x};a)|^2
d\mathbf{x}\nonumber \\
&&=-\frac{1}{8}\int\nabla^2
\left(\frac{1}{m(\mathbf{x})}\right)|\psi_{\{n\}}(\mathbf{x};a)|^2
d\mathbf{x}.
\end{eqnarray}

Let $\nabla^2 (1/m(\mathbf{x}))\leq 0$ for all $\mathbf{x}$, then
$E_{\{n\}}(a)$ is an increasing function and we get
\begin{equation}\label{eq18}
E_{\{n\}}(0)=E_{\{n\}}^{(\mathrm{BD})}<
E^{(\mathrm{LK})}_{\{n\}}=E_{\{n\}}(1).
\end{equation}
that completes the proof. For the case of
$E^{(\mathrm{LK})}_{\{n\}}$ and $E^{(\mathrm{GW})}_{\{n\}}$, the
proof is identical since the factor $\nabla^2 (1/m(\mathbf{x}))$
arises in this case as well.



It is now evident from (\ref{eq7}) and (\ref{eq16}) that if we try
to compare $E^{(\mathrm{LK})}_{\{n\}}$ with the constant-mass
energy $E^{(0)}_{\{n\}}$, the sign of the integral will be
determined by the signs of both $(1/m(\mathbf{x})-1/m_0)$ and
$\nabla^2 (1/m(\mathbf{x}))$. Unfortunately, this leads to
inconsistent conditions. For example, in order to get the
inequality $E^{(0)}_{\{n\}}>E^{(\mathrm{LK})}_{\{n\}}$, we have to
put $1/m(\mathbf{x})<1/m_0$ and $\nabla^2 (1/m(\mathbf{x}))>0$,
\textit{i.e.}, $1/m(\mathbf{x})$ must be bounded from above and
convex that is impossible. The same obstacle is encountered when
dealing with $E^{(\mathrm{GW})}_{\{n\}}$.

\section{Applications}
\label{part4}

In this section, we consider two specific PDM problems, which are
discussed in literature, and show how the comparison theorems
explain the peculiarities of their energy spectra.

\subsection{Case 1}


The three-dimensional mass distribution of the form
\begin{equation}\label{eq20}
m(r)=\frac{m_0}{(1+\kappa r)^2},
\end{equation}
with $r=|\mathbf{x}|$ and nonnegative $\kappa$, has been shown
\cite{tkachuk} to give rise to an exactly-solvable extension of
the Coulomb problem, $V(r)=-Ze^2/r$. This extension is useful as
it enables one to trace the link between the PDM background and
theories with deformations in the quantum canonical relations or
with curvature of the underlying space.

For this case, the discrete energy eigenvalues of the PDM
Hamiltonian (\ref{eq1}), in units with $\hbar=m_0=e=1$, are
written as \cite{tkachuk}
\begin{eqnarray}\label{eq21}
&&E=- \frac{\left[ Z - \frac{\kappa}{2}(l(l+1)-2\beta)
\right]^2}{2n^2}+
\frac{Z\kappa}{2} \nonumber\\
&&\phantom{E=}+\frac{\kappa^2}{8}\left[2l(l+1)-n^2-4\beta+(1+4\alpha)(1+4\gamma)
\right]
\end{eqnarray}
where $l=0,1,...$ and $n=l+1,l+2,...$ are the orbital and
principal quantum numbers, respectively. In contrast to the
constant-mass Coulomb problem, the system has only a finite number
of discrete levels, so that the allowed values of $l$ and $n$ are
restricted by
\begin{equation}\label{eq22}
\frac{\kappa}{2}[l(l+1)+n^2-2\beta]<Z.
\end{equation}
Such a restriction implies that in presence of the PDM the energy
eigenvalues may be closer to continuum and thus larger than the
ordinary Coulomb eigenenergies $E^{(0)}=-Z^2/(2n^2)$ calculated
with the mass $m_0$.

It is Theorems 1 and 2 that permit us to determine how the energy
eigenvalues are ordered. Since in (\ref{eq20}) we have $m(r)\leq
m_0$, the eigenvalues of the BenDaniel-Duke PDM Hamiltonian must
obey $E^{(\mathrm{BD})}\geq E^{(0)}$, by Theorem 1. Since
$\nabla^2(1/m(r))=(\kappa/m_0)(6\kappa+4/r)>0$, it follows from
Theorem 2 that the eigenvalues of the PDM Hamiltonians with
different ambiguity parameters are ordered as
$E^{(\mathrm{BD})}>E^{(\mathrm{LK})}>E^{(\mathrm{GW})}$.

\begin{figure}[!th]
{\parbox[!hb]{8.0cm}{
\begin{center}
\includegraphics[scale=1]{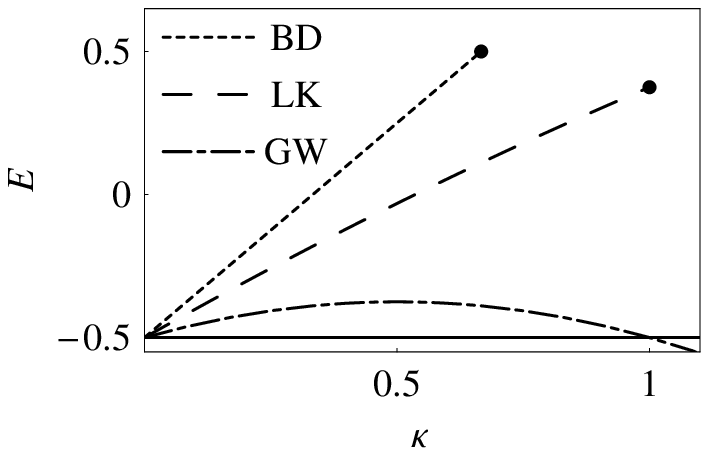}
\vspace*{-0.2cm} (a)
\end{center}}
\parbox[!hb]{8.0cm}{
\begin{center}
\includegraphics[scale=1]{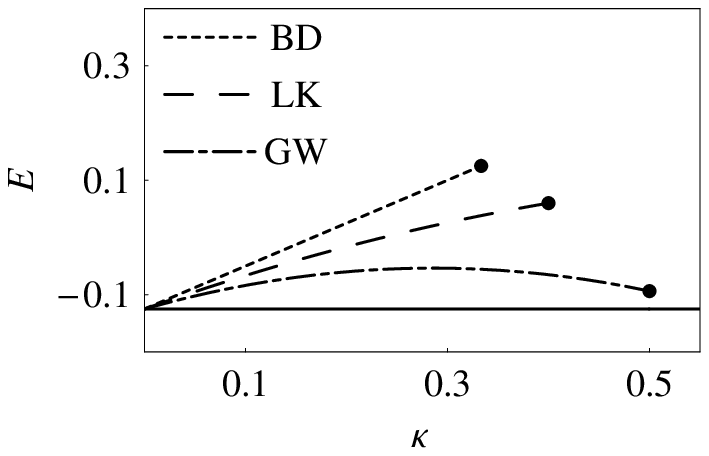}
\vspace*{-0.2cm} (b)
\end{center}}
} \vspace*{-0.2cm} \caption{Energy for the states (a) ($n=1$,
$l=0$) and (b) ($n=2$, $l=0$), in the Coulomb potential and the
mass distribution (\ref{eq20}), calculated with
$\hbar=m_0=Ze^2=1$. The solid line is the constant-mass result,
the dotted, dashed and dash-dotted curves are the PDM results
obtained with Hamiltonians (11), (12) and (13),
respectively.\hfill} \vspace*{0.25cm}

\label{Fig:1}
\end{figure}

In order to illustrate these inequalities, we present figure
\ref{Fig:1} where we plot the energy for the ground state ($n=1$,
$l=0$) and the first radially excited state ($n=2$, $l=0$), as a
function of the deforming parameter $\kappa$. In figure
\ref{Fig:1} the solid lines correspond to the constant-mass case
whereas the broken curves represent the PDM cases with different
ambiguity parameters. The circles indicate the points at which the
bound states disappear according to (\ref{eq22}). From figure
\ref{Fig:1} we see that, for all allowed $\kappa$, it holds
$E^{(\mathrm{BD})}\geq E^{(0)}$, as it was proved, and also
$E^{(\mathrm{LK})}\geq E^{(0)}$, but we observe both
$E^{(\mathrm{GW})}>E^{(0)}$ and $E^{(\mathrm{GW})}<E^{(0)}$
regions. Furthermore, we can see that the second proved
inequality,
$E^{(\mathrm{BD})}>E^{(\mathrm{LK})}>E^{(\mathrm{GW})}$, is indeed
fulfilled.

\subsection{Case 2}


Now let us consider the one-dimensional mass distribution
\begin{equation}\label{eq25}
m(x)=m_0\,(1+\kappa x^2),
\end{equation}
which is found to be useful for studying quantum wells
\cite{herling}. Applying Theorem 1 to this PDM profile, we get the
inequality $E^{(\mathrm{BD})}\leq E^{(0)}$ that justifies the
shift of electron and hole binding energies to lower values which
was observed in \cite{herling} when the spatial dependence of mass
was included. On the other hand, Theorem 2 does not apply since
the quantity $\nabla^2(1/m(x))=\kappa(6\kappa x^2-2)/m_0(1+\kappa
x^2)^3$ has an indefinite sign.

\begin{figure}[!ht]
{
\parbox[!hb]{8.0cm}{
\begin{center}
\includegraphics[scale=1]{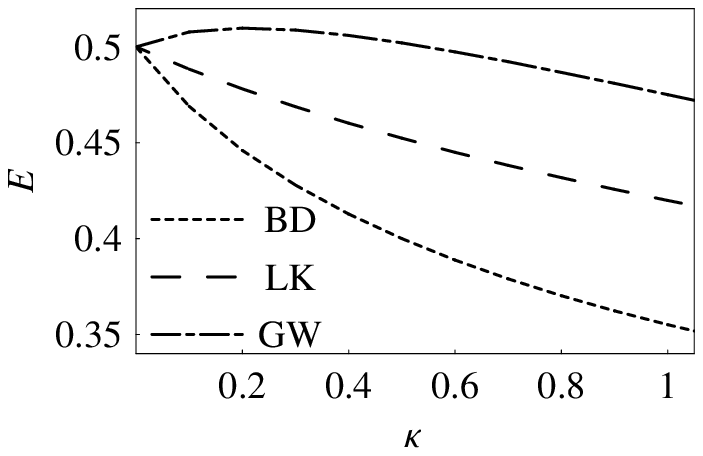}
(a)
\end{center}}
\parbox[!hb]{8.0cm}{
\begin{center}
\includegraphics[scale=1]{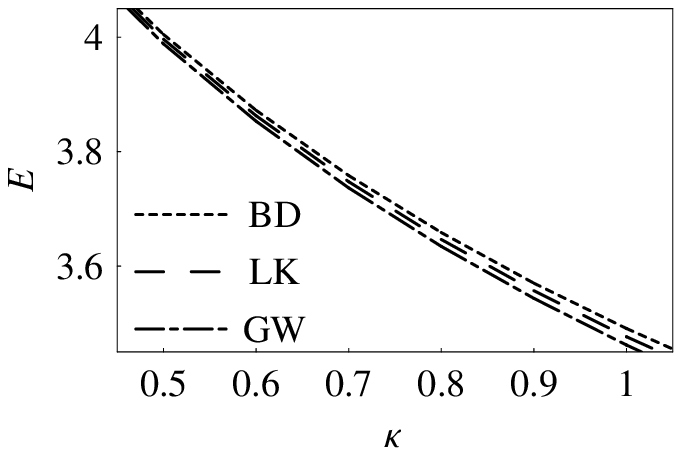}
(b)
\end{center}}
} \vspace*{-0.2cm} \caption{Energy for the states (a) $n=0$ and
(b) $n=5$, in the harmonic-oscillator potential and the mass
distribution (\ref{eq25}), calculated with $\hbar=m_0=\omega=1$.
The dotted, dashed and dash-dotted curves are the PDM results
obtained with Hamiltonians (11), (12) and (13),
respectively.\hfill} \label{Fig:2}
\end{figure}

It is worth examining how this sign indefiniteness affects the
energy spectrum. To that end, we choose the harmonic-oscillator
potential, $V(x)=\frac{1}{2}m_0\omega^2 x^2$, for which the
accurate numerical solution of the PDM Schr\"odinger equation with
the mass distribution (\ref{eq25}) is available \cite{koc2010}. In
figure \ref{Fig:2} we plot the corresponding energy of the ground
and the fifth excited states, as a function of $\kappa$, for the
three PDM Hamiltonians with different ambiguity parameters. The
energies have been calculated with $\hbar=m_0=\omega=1$, by using
the shooting method, and are in agreement with those computed in
\cite{koc2010} where the results obtained with the same $m_0$ and
$\omega$, and $\kappa=0.1$ are reported.

From figure \ref{Fig:2} it is evident that for the excited state
the discrepancy among the energies evaluated using the different
PDM Hamiltonians is less profound. However, we call attention to a
serious difference between the ground and excited states. As seen
in figure \ref{Fig:2}, the ground-state energies are ordered as
$E^{(\mathrm{BD})}<E^{(\mathrm{LK})}<E^{(\mathrm{GW})}$ whereas
the energies of the fifth excited state (and of the states with
$n>5$) are in inverse order. This inversion can be easily
understood in conjunction with Theorem 2. It is known that the
wave functions of highly excited states are spread to larger
distances. Consequently, with increasing $n$, the mean value of
$x^2$ grows and eventually reaches the point where the sign of
$\nabla^2(1/m(x))$ in Theorem 2 reverses, thus inverting the order
of energies.

\section{Summary}

\label{part5}

In this paper, we have established the comparison theorems for the
PDM Schr\"odinger equation. Our first theorem states that the
corresponding eigenvalues of a constant-mass Hamiltonian and of a
BenDaniel-Duke PDM Hamiltonian with the same potential are ordered
if the constant and position-dependent masses are ordered
everywhere. The second theorem concerns PDM Hamiltonians with the
different sets of ambiguity parameters: the BenDaniel-Duke,
Li-Kuhn, and Gora-Williams Hamiltonians. It is proved that their
corresponding eigenvalues are ordered if the Laplacian of the
inverse mass distribution $1/m(\mathbf{x})$ has a definite sign.

We have applied these theorems to the PDM Coulomb and
harmonic-oscillator problems and have been led to the following
conclusions. First, the eigenvalues of PDM Hamiltonians other than
the BenDaniel-Duke one do not have to be in the strict order with
respect to the eigenvalues of the constant-mass Hamiltonian. For
instance, from both figures \ref{Fig:1} and \ref{Fig:2} it is seen
that the order of the Gora-Williams and constant-mass ground-state
energies do vary, depending on the value of the deforming
parameter $\kappa$. Second, if the quantity
$\nabla^2(1/m(\mathbf{x}))$ has no definite sign and thus Theorem
2 does not apply, the order of the energies calculated using
different PDM Hamiltonians may alternate, as seen by comparing
parts (a) and (b) of figure \ref{Fig:2}. We therefore think that
for establishing further comparison rules within the PDM framework
one should restrict the potential profile to, \textit{e.g.}, a
spherically-symmetric case, the way the generalized comparison
theorems for the ordinary Schr\"odinger equation have been
obtained \cite{hall_gen}.

The comparison rules we have found out can be employed for
analyzing the energy spectra in semiconductor nano devices; an
example of application to the quantum well system was sketched in
the previous section. In this connection, it is worthwhile to
extend the present approach to periodic heterostructures, which
allow the direct fit of PDM binding energies to experiment
\cite{smagley}. Then we will have to abandon the requirement of
vanishing of the wave function at infinity which the proof of our
theorems relies on. What comparison rules might be formulated in
that case is an interesting open question.

\section*{Acknowledgments}

The author thanks Dr. O.Yu. Orlyansky for discussions and a
careful reading of the manuscript. The research was supported by
grant N0109U000124 from the Ministry of Education and Science of
Ukraine which is gratefully acknowledged.


\begin{thebibliography}{10}


\bibitem{bastard} Bastard G 1988 \emph{Wave Mechanics Applied to Semiconductor
Heterostructures} (Les Ulis: Editions de Physique)

\bibitem{young}  Young K 1989 \textit{Phys. Rev.} B \textbf{39} 13434


\bibitem{herling}  Herling G L and Rustgi M L 1992 \textit{J. Appl. Phys.} \textbf{71} 796


\bibitem{peter} Peter A J and Navaneethakrishnan K 2008 \textit{Physica} E \textbf{40} 2747

\bibitem{khordad} Khordad R 2010 \textit{Physica} E \textbf{42} 1503


\bibitem{dekar} Dekar L, Chetouani L and Hammann T F 1999 \textit{Phys. Rev.} A \textbf{59} 107

\bibitem{alhaidari} Alhaidari A D  2002 \textit{Phys. Rev.} A \textbf{66} 042116

\bibitem{shihai} Dong Shi-Hai, Lozada-Cassou M 2005 \textit{Phys. Lett.} A \textbf{337} 313

\bibitem{bagchi_si} Bagchi B, Banerjee A, Quesne C and Tkachuk V M 2005 \textit{J. Phys. A:
Math. Gen.} \textbf{38} 2929

\bibitem{tanaka}  Tanaka T 2006 \textit{J. Phys. A:
Math. Gen.} \textbf{39} 219

\bibitem{ganguly} Ganguly A and Nieto L M 2007 \textit{J. Phys. A: Math. Theor.} \textbf{40} 7265

\bibitem{midya} Midya B, Roy B and Roychoudhury R 2010 \textit{J. Math. Phys.} \textbf{51} 022109


\bibitem{sever}  Tezcan C and Sever R 2007 \textit{J. Math. Chem.} \textbf{42} 387

\bibitem{kraenkel} Kraenkel R A and Senthilvelan  M  2009 \textit{J. Phys. A: Math. Theor.}
\textbf{42} 415303

\bibitem{koc2010} Koc R and Sayin S 2010 \textit{J. Phys. A: Math. Theor.} \textbf{43} 455203

\bibitem{tkachuk} Quesne C and Tkachuk V M 2004 \textit{J. Phys. A: Math. Gen.} \textbf{37} 4267


\bibitem{roos} von Roos O 1983  \textit{Phys. Rev.} B. \textbf{27} 7547

\bibitem{bd} BenDaniel D J and Duke C B 1966 \textit{Phys. Rev.} \textbf{152} 683

\bibitem{lk} Li T L and Kuhn K J 1993 \textit{Phys. Rev.} B \textbf{47} 12760

\bibitem{gora} Gora T and Williams F 1969 \textit{Phys. Rev.} \textbf{177} 1179


\bibitem{morrow} Morrow R A and Brownstein K R 1984 \textit{Phys. Rev.} B \textbf{30} 678

\bibitem{mustafa1} Mustafa O and Mazharimousavi S H 2009 Phys. Lett. A \textbf{373} 325

\bibitem{almeida}  de Souza Dutra A and Almeida C A S 2000 Phys Lett. A \textbf{275} 2


\bibitem{cavalcante} Cavalcante F S A, Costa Filho R N, Ribeiro Filho J, de Almeida C A S
and Freire V N 1997 \textit{Phys. Rev.} B \textbf{55} 1326


\bibitem{smagley} Smagley V A, Mojahedi M and Osinski M 2002 {\it Physics and Simulation of Optoelectronic
Devices X} (\textit{Proc. SPIE}  vol~4646) ed P Blood, M Osinski,
Y Arakawa (San Jose: Academic) p~258




\bibitem{hall_refine} Hall R L 1992 \textit{J. Phys. A: Math. Gen.}
\textbf{25}  4459

\bibitem{hall_gen} Hall R L and Katatbeh Q D 2002 \textit{J. Phys. A: Math. Gen.}
\textbf{35} 8727

\bibitem{semay} Semay C 2011 \textit{Phys. Rev.} A \textbf{83} 024101

\bibitem{hall_rel} Hall R L 2010 \textit{Phys. Rev.} A \textbf{81} 052101

\bibitem{Hell35} Hellmann H 1935 \textit{Acta Physicochimica URSS} \textbf{1} 913;
Feynman R P 1939 \textit{Phys. Rev.} \textbf{56} 340

\bibitem{reed} Reed M and Simon B 1978 {Methods of Modern Mathematical
Physics IV: Analysis of Operators} (Academic, New York)


\end{thebibliography}
\end{document}